# Temperature-dependent non-covalent protein-protein interactions explain normal and inverted solubility in a mutant of human γD-crystallin

Amir R. Khan[1#], Susan James[2#], Michelle K. Quinn[2], Irem Altan[3], Patrick Charbonneau[3], Jennifer J. McManus[2*]

[1] School of Biochemistry and Immunology, Trinity College Dublin, Dublin 2, Ireland.

[2] Department of Chemistry, Maynooth University, Maynooth, Co. Kildare, Ireland.

[3] Department of Chemistry, Duke University, Box 90354, Durham, NC 27708-0354.

[#] authors contributed equally

[*] corresponding author, jennifer.mcmanus@mu.ie

**Author contributions:** A.K, S.J. M.K.Q, I.A. P.C and J.J.M. performed experiments and analysed data. A.K., I.A. P.C. and J.J.M wrote the manuscript.

**Abstract**

Protein crystal production is a major bottleneck for the structural characterisation of proteins. To advance beyond large-scale screening, rational strategies for protein crystallization are crucial. Understanding how chemical anisotropy (or patchiness) of the protein surface due to the variety of amino acid side chains in contact with solvent, contributes to protein-protein contact formation in the crystal lattice is a major obstacle to predicting and optimising crystallization. The relative scarcity of sophisticated theoretical models that include sufficient detail to link collective behaviour, captured in protein phase diagrams, and molecular level details, determined from high-resolution structural information is a further barrier. Here we present two crystals structures for the P23T+R36S mutant of γD-crystallin, each with opposite solubility behaviour – one melts when heated, the other when cooled. When combined with the protein phase diagram and a tailored patchy particle model we show that a single temperature dependent interaction is sufficient to stabilise the inverted solubility crystal. This contact, at the P23T substitution site, relates to a genetic cataract and reveals at a molecular level, the origin of the lowered and retrograde solubility of the protein. Our results show that the approach employed here may present an alternative strategy for the rationalization of protein crystallization.

**Introduction**

The rationalisation of protein crystallization remains a major obstacle to efficient structure determination – a requirement to understand the molecular basis for many diseases and to pinpoint targets for new drug development [1]. Sampling hundreds (or sometimes thousands) of solution conditions (i.e. mixtures of different buffers, salts and precipitants) is often the most productive strategy to identify lead conditions for protein crystallization. Even when coupled with rational design strategies, such as surface entropy reduction (SER) [2], this approach, can be time-consuming and costly since screening methods often fail to produce crystalline material or diffraction quality crystals. Protein phase diagrams which map how a given protein behaves across sets of solution conditions dramatically improve the success of the process and narrow the screening required for producing diffraction quality crystals, but have only been measured for a small number of proteins [3 and references therein]. These reference studies have identified key challenges in guiding and improving protein crystallization.

An excellent such reference is human γD-crystallin (HGD), a major structural protein found in the eye-lens. HGD is unusually stable in the eye lens in mixtures with α- and β-crystallins - often over a whole lifetime [4]. Its phase behaviour is otherwise very similar to that of a large a group of important globular proteins that includes haemoglobin [5], immunoglobulins [6], lysozyme [7], and thaumatin [8]. These phase diagrams are defined by net attractive short-range interactions that result in liquid-liquid phase separation and crystallization. Although native HGD itself does not readily form crystals, several of its genetic cataract-related single amino acid substitutions do so easily - without any major structural changes [9 - 14]. The P23T substitution which is a naturally-occurring mutation associated with congenital cataract however has unusual phase behaviour, in that its aggregates have inverted solubility, i.e. they melt as temperature is *decreased* [15, 16]. The protein is thus insoluble at physiological concentration and temperature, leading to eye lens opacity. In the related P23V mutant, both aggregates and crystals are observed, both also with inverted solubility [15, 16]. Yet, crystallization of the P23T mutant under physiological conditions has remained elusive. Numerous structural and biophysical studies, including X-ray structures at pH 4.5 and NMR solution studies, have failed to unambiguously identify major structural changes in the P23T mutant and hence a full explanation for its anomalous inverted solubility is still unknown [17 - 20].

Physico-chemical insights into protein phase behaviour – both normal and anomalous, are often gleaned from colloidal science. Simple colloidal models do capture key features of protein phase diagrams, such as their metastable critical point [21]. However, protein phase diagrams cannot be completely rationalised without including some level of anisotropy, in terms of the directional contacts between proteins in solution or within a crystal lattice [22 - 28] or of shape anisotropy [29]. This anisotropy gives rise to rich protein phase diagrams and is more widely exploited for the controlled assembly of biological and biomimetic materials [30]. It has even been proposed that these types of interactions are important in controlling liquid-liquid phase separation in cells [31], with important implications in understanding stress responses, RNA processing and gene expression.  However, understanding and predicting anisotropic protein-protein interactions *ab initio* is not yet possible, due to the extreme heterogeneity of amino acid side chains on the protein surface. While measurements indicative of net protein-protein interactions, such as the osmotic second virial coefficient, $B_{22}$, or the diffusivity constant, $k_D$, can provide some insight, they reflect the averaged pair interactions between proteins. These parameters are typically insufficient to trace back the specific, directional protein-protein interactions that control the dramatic (and often unpredictable) changes in protein assembly upon mutagenesis [14, 16]. Enhanced numerical models, that capture the details of anisotropic protein-protein interactions may allow for the prediction of protein phase diagrams and hence optimal crystallization conditions [17, 21-29, 32-35]. To identify the microscopic origin of inverted solubility, however, we need high-resolution structural information detailing the underlying anisotropic interactions, using for instance crystal structures of the protein of interest.

To design a P23T mutant that crystallizes at pH 7, we focussed our interest on HGD mutant structures that do not form specific protein –protein contacts near proline 23. One such mutant, R36S readily crystallizes by forming a crystal lattice contact at position 36. By combining the R36S and P23T substitutions we reasoned that crystals of the double mutant would display inverted solubility, based on a comparison of the phase diagrams for the single mutant proteins, thus providing insights into the mechanism for the P23T mutant retrograde solubility. Remarkably, the double mutant, P23T+R36S, formed two distinct crystals forms - one with normal solubility and one with inverted solubility [36]. While inverted solubility in proteins has been previously observed, a protein that forms two distinct crystal lattices, each with opposite temperature dependence of the solubility line, had not and therefore this double mutant offers a rare opportunity to access the microscopic origin of solubility inversion, which we now probe further.

Here we report the X-ray structures of the two crystal forms of the P23T+R36S mutant of HGD. We find that the two are polymorphs with different unit cells and crystal contacts and that it is possible to interchange between them by solely varying the solution temperature. In the inverted solubility crystal, a lattice contact involving the cataract-associated Thr23 residue is formed. This is a new contact with the same binding energy determined from a statistical mechanics analysis of the chemical potentials of the solubility lines in earlier work [16]. We have used both the phase diagram for P23T+R36S and crystal structures to design a custom patchy particle model which incorporates specific contacts formed in the crystal lattice. We find that when temperature dependent patchy interactions are included, the temperature dependence of the solubility lines for both crystal lattices can be reproduced by simulations performed using the custom model. Specifically, we show that a change to the contact that contains the 23$^{rd}$ residue in the inverted solubility crystal is sufficient to cause inverted solubility. This contact becomes engaged as temperature increases, stabilising the inverted solubility crystal phase, thus revealing the molecular origin of the inverted solubility for P23T.

**Results and Discussion**

The equilibrium phase diagram for P23T+R36S is shown in figure 1 [36]. Two different crystal types are observed, distinguished by the temperature-dependence of their respective solubility lines; one with normal solubility (DBN), which melts as temperature increases and a second with inverted solubility (DBI), which forms at higher temperatures and melts as temperature is lowered. The solubility lines intersect at ≈ 303 K, where both crystal forms coexist. Remarkably, the two crystals form under near physiological conditions of temperature, pH and salt, unlike the previously determined structure of P23T (pH 4.6, PEG4K) [20].

The crystals have different morphologies; DBN crystals are rod shaped and DBI crystals are rhombic. The proteins remain in their fully-folded globular state across the temperatures probed in this work. We further showed in previous work that this mutant protein displays no significant change to its secondary structure relative to native HGD. Our determination of the structures for these two crystal forms by X-ray crystallography confirms this finding for these polymorphs.

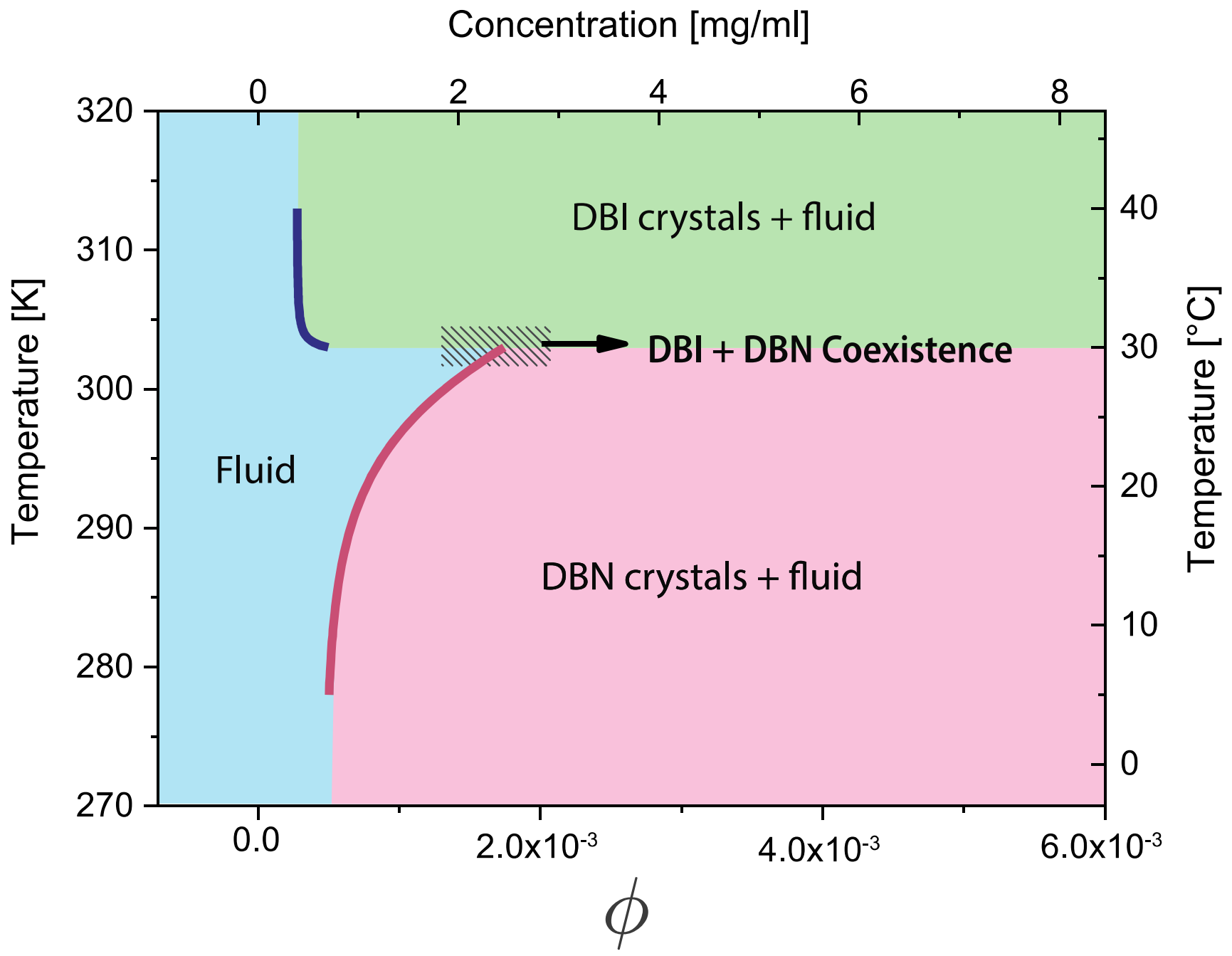

*Figure 1: Phase Diagram for P23T+R36S mutant of human γD-crystallin, indicating the equilibrium phase boundaries for the two crystals formed and their respective fluid phases. The volume fraction ($\phi$) is calculated as $\phi = c.\upsilon_{sp}$, where c is the concentration of protein in mg/ml and* $v_{sp}$ = 7.1 x $10^{-4}$ mg/ml [16]. *Coexistence of the two crystals is observed the temperature at which the phase boundaries overlap (~303K).*

The DBI (PDB ID: 6ETC) and DBN (PDB ID: 6ETA) structures consist of paired homologous domains that each adopt a Greek key motif. DBI crystallized as a monomer at high resolution (1.2 Å), while DBN crystallized with two molecules in the asymmetric unit at medium resolution (2.2 Å).

The overall structures of DBN and DBI are otherwise highly conserved, with an RMSD of 0.44 Å for the main chain atoms (residues 1-173). The side chain of Thr23 in DBI is involved in a crystal contact, within which it forms a hydrogen bond to the backbone of Gly129 in a symmetry-related molecule (figure 2a). This interaction is unique to DBI as Thr23 is not involved in lattice contacts of DBN. These polar interactions unambiguously demonstrate that the pathogenic P23T mutation enables direct interactions in the crystal lattice. By contrast, Ser 36 is not involved in any direct contact in the DBI crystal (figure 2b). It only contributes a hydrogen bond within a DBN crystal contact.

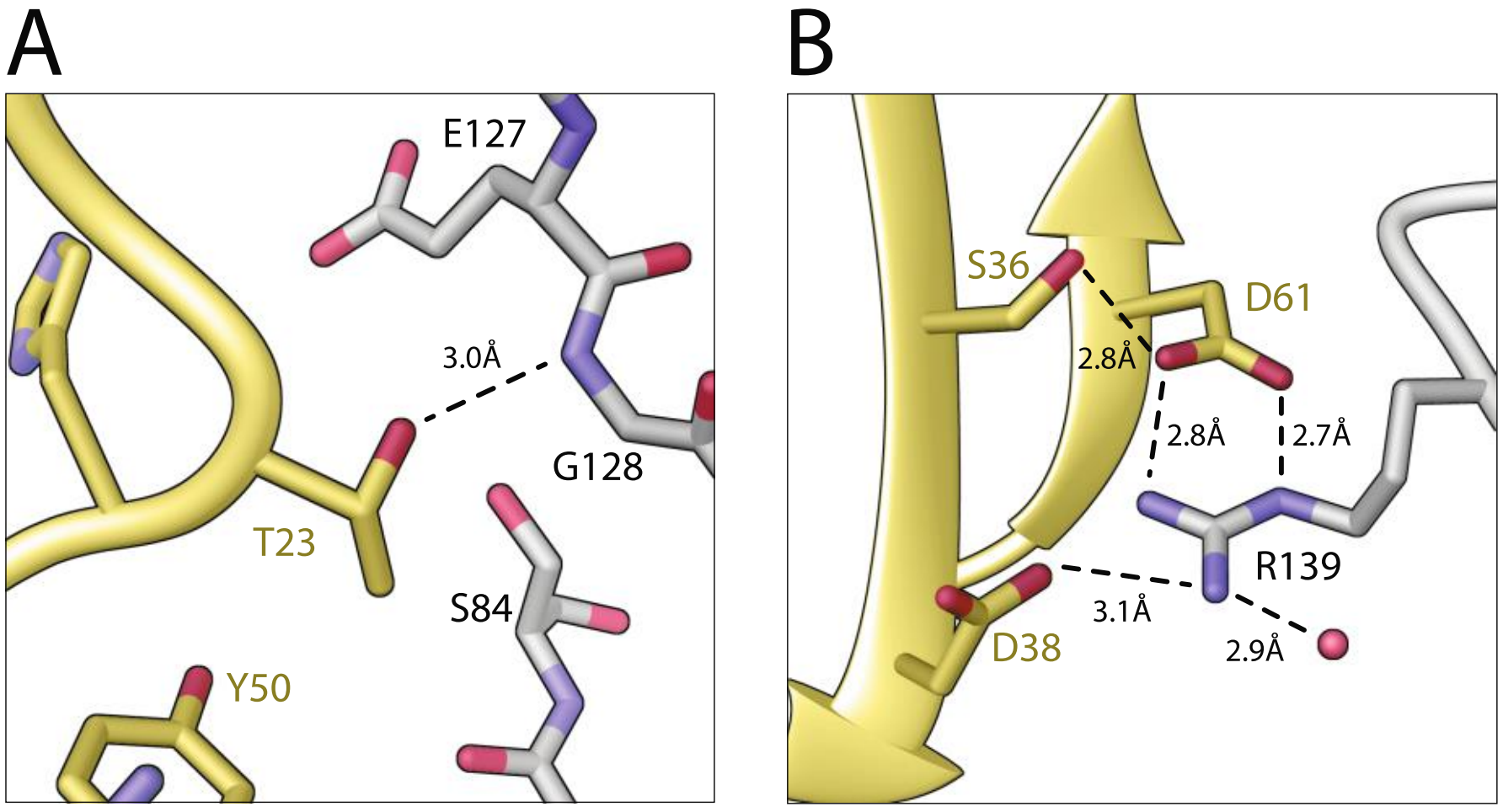

*Figure 2: (a) Interactions between Thr23 and the crystal lattice. The side chain of Thr23 (gold) forms a hydrogen bond with the backbone of NH of Gly128, indicating the close contacts between the P23T locus and a symmetry-related molecule; (b) Crystal contacts near the R36S locus of DBI. Ser36 does not make direct contacts in the crystal lattice. However, the side of Ser36 makes a hydrogen bond with Asp61, which makes salt bridges Arg139 in a symmetry related molecule.*

If we are to relate our findings to the P23T single mutant, it is reasonable to ponder if the P23T+R36S mutant is a good model for it. The R36S contact is not activated in DBI suggesting that it does not influence the structure of the DBI crystal to any significant extent and the DBN crystal is has the same structure and lattice contacts as the R36S single mutant. Because P23T and R36S reside on opposite sides of the N-terminal domain, we expect the structural and energetic influence of the two to be uncoupled. The molecular interactions at the 36 locus are also distinct. In the structures of DBN and the R36S single mutant (PDB ID: 2G98) [12], Ser36 forms a hydrogen bond with Asn24 with a symmetry mate in the lattice. In contrast, DBI forms an intramolecular hydrogen bond with Asp62, which in turn ion-pairs with Arg140 in a symmetry-related molecule. Thus, Ser36 in DBI orients Asp62 lattice interactions, which is distinct from DBN.

There is a significant degree of flexibility in the C-terminal domain of DBN (molecule B), which likely explains why only a medium resolution structure could be obtained as evidenced by the associated B-factors (Table 1). Other HGD mutants with medium resolution structures, namely the P23T (PDB ID: 4JGF) [20] and R36S (PDB ID: 4JGF) [12] single mutants, indeed display comparable flexibility in the C-terminal domain. By contrast, the corresponding domain in the DBI crystal is more rigid. It has a crystal contact formed by Ser173 (Oγ) and the backbone NH of Gly157 of a symmetry-related molecule (figure 3A), which is associated with the higher-resolution structure. Strikingly, the C-terminal carboxylate forms an ion-pair with Arg141 from the same symmetry mate. There are also non-polar interactions between Phe172 and Gln67 from a second symmetry-related molecule, indicative of the intimate associations between the C-terminus of DBI in the crystal lattice. It is not clear, however, whether this flexibility is a result of a lack of a stabilizing crystal contact, or whether, conversely it precludes contact formation.

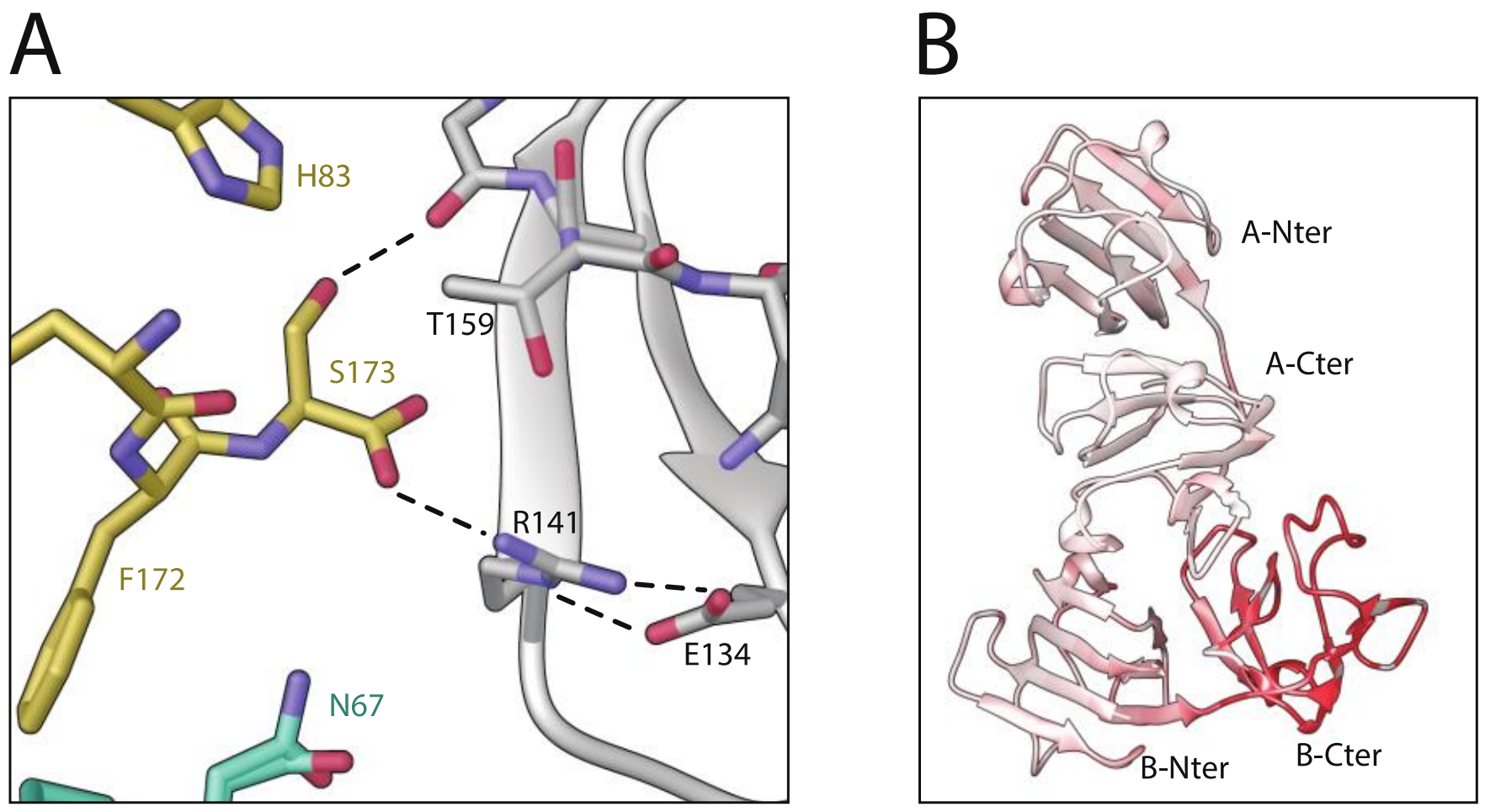

*Figure 3: (A) Interactions at the C-terminus of DBI (gold). DBI is green and symmetry-related molecules are grey and teal, respectively. (B) Flexibility of the C-terminal domain of DBN. The two ribbon models (A and B) in the asymmetric unit are annotated by gradient colours of backbone B-factors. Region with high values indicating flexibility are red, while ordered regions are white.*

The formation of a hydrogen bond between Thr23 and the backbone of Gly129 in DBI suggests a molecular basis for a change in protein-protein interactions in the region of position 23 in the mutant protein. The change in the net binding energy between native HGD and P23T single mutant protein calculated from the solubility data [16], corresponds to approximately 2.4 $k_BT$, which is indeed the strength of a typical hydrogen bond. Beyond this observation, there is no obvious structural basis for the inverted temperature dependence of the solubility line. Therefore, we employed a modelling strategy, based on custom patchy particle colloidal models to investigate the microscopic origins of the inverted solubility of the double mutant.

The model describes proteins as having a hard spherical core with directional, short-ranged attractive patches representing crystal contacts, derived from the crystal structures (see SI). DBI and DBN are modelled with five contacts each (figure 4, see SI for details, including the amino acids involved in the different contacts). Despite its very crude description of protein-protein interactions, such models can recapitulate the characteristic topology of protein phase diagrams. Because solubility inversion necessarily implies some degree of temperature dependence for the patch interactions [36], we first consider deactivating the contact that contains the 23$^{rd}$ residue, where the new crystal contact is formed, around a temperature $T_a$ with rate $\tau$, set by the experimentally observed inverted solubility temperature and density ranges, respectively.

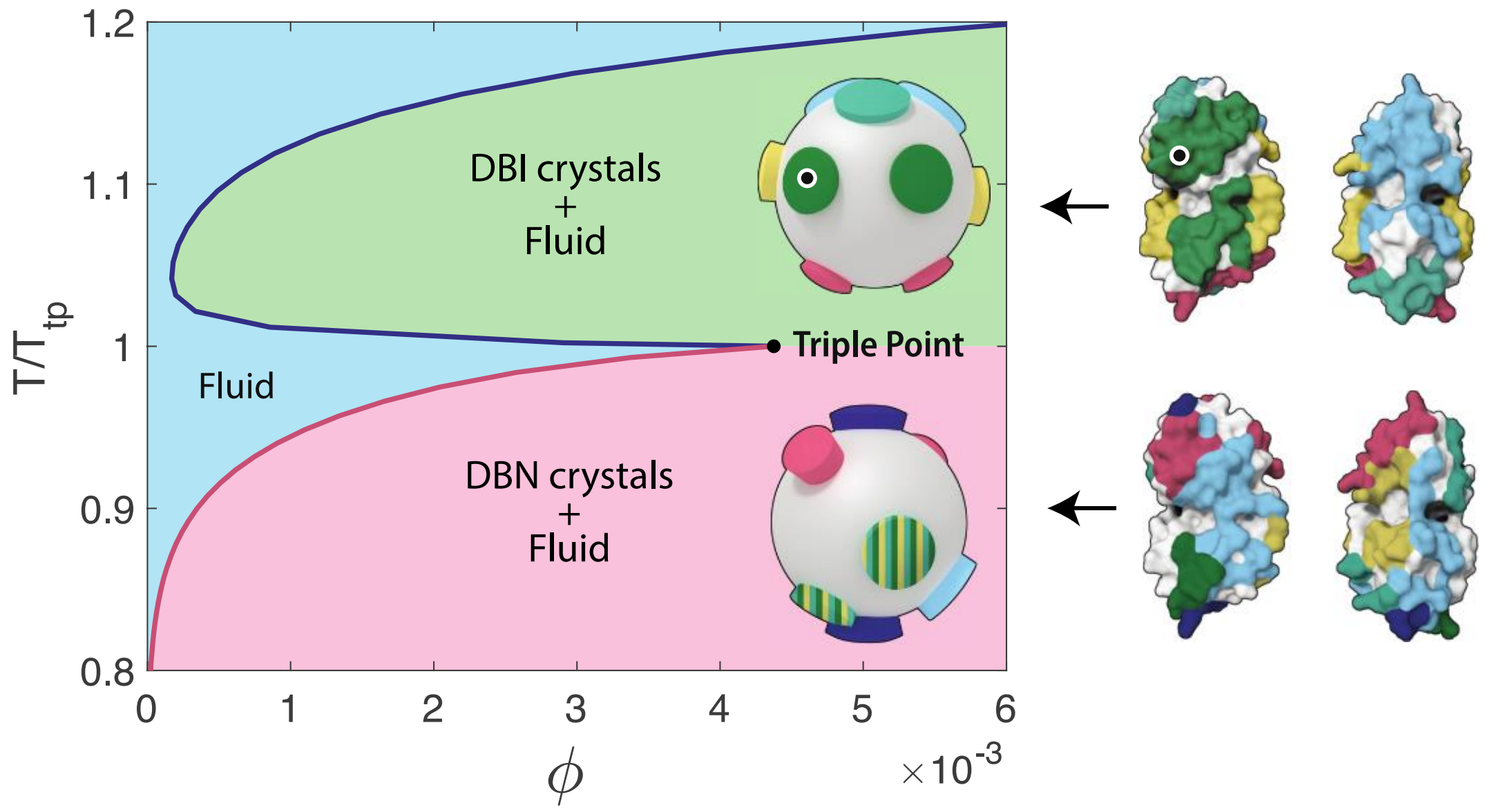

*Figure 4: Crystal structures of DBI and DBN (right) are used to devise patchy particle models (centre). Patches derive from crystal contacts and each of them is represented here by a different colour (see table S1). A black dot specifically denotes the mutated 23$^{rd}$ residue. The resulting DBI (red) and DBN (blue) solubility lines obtained by deactivating the contact containing the 23$^{rd}$ residue below temperature $T_a$ intersect at the nearby triple point temperature, $T_{tp}$, in near quantitative agreement with experimental observations.*

Simulations of this model with specialized Monte-Carlo methods determined the equilibrium phase diagram shown in figure 4. Upon cooling, the model solubility line for DBI crystals reaches a minimal volume fraction, $\phi{\sim}10^{-4}$, before exhibiting an inverted solubility behaviour, and then the DBI crystal phase quickly disappears, all in remarkable agreement with experimental observations. The DBN solubility line, which shows normal solubility, afterwards intersects with that of the DBI crystal around $\phi{\sim}10^{-3}$, forming a triple point. Experimental results are also suggestive of a triple point for comparable densities, but the flatness of the DBI solubility line in this regime precludes its accurate determination. Note that although similarly deactivating a larger set of DBI patches can also reproduce the observed experimental phase behaviour, no microscopic basis exists for these changes, and doing so to more than a couple of patches melts the crystal before solubility inversion can be observed. Prior experimental observations suggest that a change to surface hydrophobicity using either small molecule dyes [32] or by mutagenesis at position 23 [37] may give rise to entropic gain upon crystallization and could explain the lowered solubility of the mutant protein. Our results are consistent with this proposal.

In our model, although the crystal state energy of DBI is significantly lower than that of DBN, this difference is largely offset in the inversion regime by the pressure-volume contribution that arises from the packing fraction difference between the two crystal forms and from the entropy difference associated with the different constraints on the crystal forms. The resulting chemical potential difference between the two crystal forms is then sufficiently small for DBI to become more stable than DBN upon activating a single patch. The solubility is inverted simply because activating the crystal contact that includes the 23rd residue substantially increases the stability of the DBI crystal. This analysis thus confirms that local changes around the mutated residue alone can give rise to the observed inverted solubility, and that few other explanations are physically reasonable.

## Methods

*Preparation and characterisation of double mutant*

The double mutant was created, expressed and purified as described previously [36]. SDS-PAGE and size exclusion HPLC were used to confirm protein purity at >98%. The intact molecular weight for the mutant protein was analysed by electrospray ionization mass spectrometry (Finger Prints Proteomics Facility, College of Life Sciences, University of Dundee), which confirmed molecular mass of 20,541 ±1 Da for P23T+R36S mutant.

*Crystallization and data collection*

The crystals were from the P23TR36S double mutant of human γD-crystallin protein were obtained and grown in grown in capillaries in 100mM Sodium Phosphate buffer (pH 7) in the presence of 20mM DTT. The solution concentration of protein was in the range of 1-2 mg/ml and there was no additional precipitant in the solution. Crystals of the double mutant with inverted solubility (DBI) were grown at 310K, while crystals with normal solubility (DBN) grew at 277K. Crystals were harvested from capillaries and mixed with 25% glycerol, flash-cooled in liquid nitrogen and subjected to X-ray diffraction. Data were collected at the PX2 beamline at Le Soleil Synchrotron, on an ADSC Q315 detector.

*Solubility measurements*

Protein solutions were prepared initially by diafiltration against 100mM sodium phosphate buffer, pH 7.0 using Ultracel 10KDa ultrafiltration discs (Merck Millipore, Co.Cork, Ireland). Protein concentrations for the double mutant was measured by UV absorbance using the extinction coefficient value of 2.09 $mg^{-1}$ ml $cm^{-1}$ after filtration through 0.22µm Millex-GV Millipore (Merck Millipore, Co.Cork, Ireland) syringe driven filters. When required, protein solutions were further concentrated by ultrafiltration using Amicon Ultra-4 centrifugal filter units (Merck Millipore, Co.Cork, Ireland) and the protein concentration re-established by UV absorbance.

*Data processing and structure solution*

The structure of DBN was solved using the model that contains the R36S single-site mutation in human γD-crystallin (PDB ID: 2G98) [38]. The program Phaser [39] provided the starting model, which was improved through cycles of manual model-building using Coot [40] and Phenix refinement [41]. The structure of DBI was solved using the high-resolution 1.25Å structure of wild-type human γD-crystallin (PDB ID: 1hk0 [13]). The obtained structure was further refined using the same refinement procedure as in DBN. Statistics from the data collection and refinement strategies are detailed in Table 1. Crystal contacts determined from the structural analysis were used to determine the patch-patch interactions for the phase diagram, as described below.

*Description of the Model*

Since transitions between the two crystal forms occur upon temperature change, we consider the phase behaviour of the double mutant using a patchy particle model with temperature-dependent patches. This choice accounts for the associated change in bonding free energy [16].

*Model Definition*

Proteins are modelled as patchy particles with interactions adapted from the Kern-Frenkel model [42]. Hard spheres (HS) with a diameter σ, chosen as the largest centre of mass distance between protein-protein crystal contacts, interact with directional, attractive patches of range $\lambda_{\alpha\beta}\sigma$ for each protein-protein contact $\alpha\beta$. The patch interaction potential,

$$u(r_{ij}, \Omega_i, \Omega_j) = u_{HS} + \sum_{\alpha,\beta}^{n} u_{\alpha\beta}(r_{ij}, \Omega_i, \Omega_j),$$

thus includes a factorized attractive contribution, $u_{\alpha\beta} = v_{\alpha\beta}(r_{ij}) f_{\alpha\beta}(\Omega_i, \Omega_j)$, that depends on interparticle distance, $r_{ij}$, and particle orientations, $\Omega_i$ and $\Omega_j$. Its orientational component is

$$f_{\alpha\beta} = \begin{cases} 1, & \theta_{\alpha,ij} \leq \delta_\alpha \text{ and } \theta_{\beta,ij} \leq \delta_\beta \\ 0, & \text{otherwise} \end{cases} \times \begin{cases} 1, & \psi_{\alpha\beta,ij} \in [\phi_{\alpha\beta} - \Delta\phi_{\alpha\beta}, \phi_{\alpha\beta} + \Delta\phi_{\alpha\beta}] \\ 0, & \text{otherwise} \end{cases}$$

where the first term ensures that patch vectors are facing each other (SI Fig. 1). The second restricts the torsion between the two particles (SI Fig. 2). Its radial component is a square-well potential

$$v_{\alpha\beta}(r_{ij}) = \begin{cases} -\varepsilon_{\alpha\beta}(T), & \sigma < r_{ij} < \lambda\sigma \\ 0, & \text{otherwise} \end{cases}$$

where $\varepsilon_{\alpha\beta}$ is constant if the patch is not temperature-dependent, and otherwise has a modulated interaction: $\varepsilon_{\alpha\beta}(T) = \frac{\tilde{\varepsilon}_{\alpha\beta}}{2}(1 + \tanh(^{T - T_a}/_{\tau_{\alpha\beta}}))$, [43] which becomes deactivated below temperature $T_a$ over a rate set by $\tau_{\alpha\beta}$, thus capturing the change in free energy due to the effect of proposed increase in hydrophobicity. Model parameters were determined from all-atom molecular dynamics simulations of all patches of DBI and DBN. (See SI for methodological details and model parameters.)

*Phase Diagram Determination*

The phase diagram of the schematic model is obtained by specialized Monte Carlo (MC) simulations: (i) the reference crystal free energies are obtained by integrating from an ideal Einstein crystal using the Frenkel-Ladd method [44]; (ii) the chemical potentials of the crystal phases as a function of temperature are obtained by thermodynamic integration along isobars from the reference in (i); (iii) fluid free energies are approximated using the second virial coefficient, $B_{22}$, due to inefficiency of traditional MC sampling at low fluid densities (SI). Coexistence points between the fluid and crystal phases are determined from the intersection of chemical potential curves, and coexistence lines then are traced out using a Gibbs-Duhem integration scheme [45, 46].

## Conclusions

The rational design of a double mutant based on phase diagrams of single mutant proteins has allowed us to produce two crystal forms of the P23T+R36S mutant of HGD, that are polymorphs with different unit cells and distinct crystal contacts. The use of a single amino acid substitution (R36S) previously shown to increase the crystallization propensity of HGD but which is unrelated to the mutant under consideration (P23T) is not standard but could provide an alternative design strategy to assist large-scale crystallization screening. The crystal displaying inverted solubility (DBI) forms a hydrogen bond at position 23, which distinguishes it from other gamma crystallin structures. Employing crystallographic data for both crystals, further investigation of the microscopic origin of inverted solubility and greater understanding of the solution behaviour of the P23T single mutant was possible. By considering a patchy particle model parameterized for this particular system, the phase diagram for the double mutant protein was reproduced by simulations. A single temperature-dependent contact, specifically the contact that includes the P23T mutation, is sufficient to explain the crystallization behaviour for the protein. Activation of the patch that contains this mutation was found to stabilize the inverted solubility crystal. This overall analysis illustrates that although non-covalent protein-protein interactions are far from trivial and thus challenging to predict, the combination model and experimental phase diagrams could be a productive approach to rationalize and provide support for future crystallization studies.

**Acknowledgements**

This work was made possible by Science Foundation Grant 11/RFP.1/PHY/3165, the Irish Research Council and SFI Stokes Lectureship to JJMcM. The authors thank Nicolette Lubsen for permission to use the HGD plasmid DNA. IA and PC thank Diano Fusco for supporting discussions and acknowledge support from National Science Foundation Grant No. DMR-1749374. Computational parts of this work used the Extreme Science and Engineering Discovery Environment (XSEDE), which is supported by National Science Foundation Grant no. ACI-1548562, as well as the Open Science Grid and the Duke Compute Cluster. ARK was supported by Science Foundation Ireland (grant number 12/IA/1239). The authors wish to thank beamline scientists at Proxima 2 (Soleil Synchrotron) for their help in data collection.

**Table 1. Data collection and refinement statistics.**

Statistics for the highest-resolution shell are shown in parentheses.

**B*-factors for the C-terminal domain of molecule B reveal significant domain flexibility

| | **DBI** | **DBN** |
|---|---|---|
| **Resolution range** | 44.01-1.197 (1.24 - 1.197) | 48.16 – 2.20 (2.277-2.20) |
| **Space group** | P 1 21 1 | P 21 21 21 |
| **Unit cell (Å)** | 44.02 31.70 52.50 90 91.29 90 | 54.04 82.10 106.25 90 90 90 |
| **Total reflections** | 89029 (8455) | 164792 |
| **Unique reflections** | 45768 (4504) | 24656 (2391) |
| **Multiplicity** | 1.9 (1.9) | 6.7 |
| **Completeness (%)** | 99.46 (99.03) | 99.63 (98.03 |
| **Mean I/sigma(I)** | 8.93 (1.22) | 9.2 (1.4) |
| **Wilson B-factor** | 10.73 | 38.69 |
| **R-merge** | 0.03956 (0.5534) | 0.129 |
| **R-meas** | 0.05595 (0.7827) | 0.139 |
| **R-pim** | 0.03956 (0.5534) | 0.053 |
| **Reflections used for R-free** | 1999 (190) | 1233 (119) |
| **R-work** | 0.1493 (0.2662) | 0.2304 (0.2993) |
| **R-free** | 0.1794 (0.2835) | 0.2658 (0.3614) |
| **Number of non-hydrogen atoms** | 1687 | 2888 |
| **macromolecules** | 1479 | 2776 |
| **solvent** | 208 | 112 |
| **Protein residues** | 173 | 341 |
| **RMS(bonds)** | 0.013 | 0.025 |
| **RMS(angles)** | 1.21 | 1.49 |
| **Ramachandran favored (%)** | 98.83 | 94.93 |
| **Ramachandran allowed (%)** | 1.17 | 4.18 |
| **Ramachandran outliers (%)** | 0.00 | 0.9 |
| **Rotamer outliers (%)** | 1.85 | 0.69 |
| **Average B-factor** | 14.98 | 56.9 |
| **macromolecules** | 13.30 | 57.26 |
| **solvent** | 26.91 | 47.89 |
| **chain A (DBN): 1-81** | --- | 40.16 |
| **chain A (DBN): 82-173** | --- | 40.81 |
| **chain B (DBN): 1-81** | --- | 51.48 |
| **chain B (DBN): 82-173** | --- | *101.38 |
| | | |

# Supplementary Information

## Temperature-dependent non-covalent protein-protein interactions explain normal and inverted solubility in a mutant of human γD-crystallin

Amir R. Khan[1#], Susan James[2#], Michelle K. Quinn[2], Irem Altan[3], Patrick Charbonneau[3], Jennifer J. McManus[2*]

[1] School of Biochemistry and Immunology, Trinity College Dublin, Dublin 2, Ireland.
[2] Department of Chemistry, Maynooth University, Maynooth, Co. Kildare, Ireland.
[3] Department of Chemistry, Duke University, Box 90354, Durham, NC 27708-0354.
[#] authors contributed equally
[*] corresponding author, jennifer.mcmanus@mu.ie

### Contacts and Amino Acids in Each Contact

The amino acids involved in DBI crystal contacts are given in table S2 and those for DBN can be found in table S3. The mapping of these amino acids on the protein surface are shown for different contacts in figure 4 of the manuscript.

*Table S2 List of amino acids in DBI crystal contacts. The residues shown in bold are involved in a hydrogen bond or a salt bridge, as identified by the PISA server [13]. Mutated residues are marked with an asterisk. The colors in parentheses are key to figure 4 in the manuscript.*

| DBI contacts | Amino acids within 5A of interface | | |
|---|---|---|---|
| c1 (light blue) | **ASP38** | LEU135 | **GLY140** |
| | **VAL37** | SER36* | **GLY10** |
| | LEU71 | HIS65 | **ARG9** |
| | SER173 | **ASP64** | **ARG167** |
| | **PHE11** | GLY60 | **ARG162** |
| | **GLN66** | **ASP61** | ARG168 |
| | GLN67 | ALA63 | ARG141 |
| | MET69 | **ARG58** | LEU53 |
| | TYR138 | **ARG59** | **TYR62** |
| | **ARG139** | PHE172 | **ASP107** |
| | GLU134 | GLN12 | |
| c2 (teal) | **SER30** | **TYR28** | ASP113 |
| | **GLN100** | ASN118 | **GLN112** |
| | **GLU7** | ARG114 | **ARG14** |
| | **PHE115** | PHE117 | **PRO27** |
| | LEU29 | **ARG116** | |
| c3 (gold) | TYR153 | **ARG59** | GLU106 |
| | GLN47 | ASN160 | **GLU127** |
| | **ARG168** | HIS87 | SER86 |
| | GLY157 | HIS83 | ARG141 |
| | **THR159** | SER39 | SER84 |
| | GLY85 | ALA158 | LEU53 |
| | ILE170 | TYR143 | **GLU46** |
| | **GLN154** | PHE172 | GLY52 |
| | **ARG152** | SER173 | |
| c4 (green) | ASP155 | MET146 | GLY148 |
| | TYR150 | HIS83 | GLU127 |
| | ARG88 | SER20 | PRO27 |
| | **ASN49** | **HIS22** | **ASP149** |
| | TYR97 | ASP21 | SER86 |
| | GLN154 | **ARG151** | SER84 |
| | TYR16 | **GLY128** | TYR50 |
| | ARG152 | SER129 | TYR28 |
| | **ASN24** | PRO147 | **THR23*** |
| | **VAL125** | PRO48 | PRO82 |
| | **GLU17** | **LEU126** | |
| | SER19 | **LYS2** | |
| c5 (red) | PHE104 | **ARG88** | TYR92 |
| | **GLU103** | **ARG98** | ARG94 |
| | **GLY99** | MET101 | **GLU93** |
| | **THR105** | **ARG90** | |

*Table S3 List of amino acids in DBN crystal contacts. The residues shown in bold are involved in a hydrogen bond or a salt bridge, as identified by the PISA server. Mutated residues are marked with an asterisk. The colors in parentheses are key to figure 4 in the manuscript.*

| DBN contacts | Amino acids within 5A of interface | | |
|---|---|---|---|
| cI (dark blue) | **ARG116** | GLN12 | ARG31 |
| | GLY13 | GLY99 | **ASP8** |
| | **GLN100** | ARG98 | |
| | **GLU7** | ARG14 | |
| cII (gold) | ARG114 | **ASN137** | SER109 |
| | ARG88 | **ASP113** | ARG167 |
| | **ARG116** | GLY60 | GLU106 |
| | GLN112 | **ASP61** | **GLU103** |
| | **GLN67** | HIS87 | GLU127 |
| | **GLY85** | **ARG58** | SER86 |
| | TYR138 | ARG59 | TYR62 |
| | ARG139 | THR105 | ASP107 |
| cIII (light blue) | **TYR150** | LEU135 | PRO147 |
| | LEU71 | **TYR50** | LEU126 |
| | ARG88 | SER19 | **ASP113** |
| | **ARG116** | MET146 | GLY148 |
| | GLN112 | **LEU80** | **ARG162** |
| | GLN66 | ARG79 | GLU127 |
| | MET69 | THR23* | ASP149 |
| | CYS110 | HIS22 | **HIS83** |
| | TYR97 | **ASP21** | **TYR28** |
| | TYR16 | ASN124 | SER20 |
| | **ARG139** | PHE117 | PRO82 |
| | GLU134 | **ASN118** | GLY140 |
| cIV (green) | ARG151 | **ASP149** | **ASP96** |
| | **ASP155** | GLY99 | **ARG94** |
| | GLN100 | **ARG98** | GLU93 |
| cV (red) | GLY10 | **ASN24** | **PHE11** |
| | ALA63 | **GLN12** | PRO27 |
| | **GLN47** | ASN49 | **HIS22** |
| | ASP38 | **ASP61** | **GLN26** |
| | **TYR6** | **SER36*** | LYS2 |
| | ARG59 | TYR28 | PRO48 |
| | THR23* | THR4 | |
| cVI (teal) | ARG114 | ASP73 | ARG31 |
| | **GLN47** | **ASP113** | GLU106 |
| | LEU71 | GLU46 | |
| | THR105 | SER72 | |
| cVII (red) | GLY10 | **ASN24** | LYS2 |
| | ARG9 | **GLN12** | **SER36*** |
| | **GLN47** | PRO48 | ALA63 |
| | **GLN26** | TYR28 | THR4 |
| | **TYR6** | ASN49 | THR23* |
| | PHE11 | PRO27 | |

| | | |
|---|---|---|
| | ASP61 | ASP38 |

## Calculation of Model Parameters

Model parameters are calculated from all-atom molecular dynamics (MD) simulations using Gromacs [1] (versions 5.1 and 2016). As input configuration, pairs of proteins were placed in their crystal contact conformation. The missing atoms and residues to the crystal structures were inserted and the resulting structure was placed in a simulation box along with water and salt. Because no force fields exist for sodium phosphate, sodium chloride in identical ionic strength was included. The key impact of this salt on protein-protein interactions is to tune the ionic strength, hence the net result is equivalent. In addition, DTT was not included because its dominant role is to prevent disulfide bonds from forming, which is in any case impossible in our classical MD simulations.  The simulation box was then relaxed by energy minimization followed by a short, 100 ps NPT simulation. Simulation parameters that are common to the various types of MD simulations are given in table S4.

In order to parameterize the patchy model, the potential of mean force (PMF) was obtained for each crystal contact. by running umbrella sampling simulations. These simulations were prepared by first pulling one of the protein copies away from the reference protein, following a direction perpendicular to the contact interface. Meanwhile, the reference protein is kept fixed by restraining the positions of three or four alpha carbons, chosen such that they are approximately equally-spaced, are not coplanar, and are not too close to the contact interface. To prevent the pulled protein from rotating, *x*- and *y*-coordinates of the same set of alpha carbons were also restrained. Pulling is done with a harmonic spring with a spring constant of $5000\ \mathrm{kJ\ mol^{-1}nm^{-2}}$, at a rate of 0.01 nm per ps. Input configurations for the umbrella sampling are then generated from the resulting trajectory, and then relaxed with a 100 ps NPT simulation with the same harmonic constraints as in the pull simulation, before running 20 ns-long trajectories. Force information is saved every 100 ps to generate the PMF with Weighted Histogram Analysis Method implemented within Gromacs [2].

The parameters for each patch were then determined as follows. Its square-well potential depth was chosen as the depth of the PMF from the well to the long-distance plateau. Its interaction range was calculated by matching its contribution to the second virial coefficient, $B_{22,\alpha\beta}$, to that of the PMF hence

$$B_{22,\alpha\beta} = -\frac{1}{2}4\pi\int\left(e^{-\beta U(r)}-1\right)r^2dr.$$

The interaction range is then found by

$$\lambda_{\alpha\beta} = \left(\frac{3\int\left(e^{-\beta U_{PMF,\alpha\beta}(r)}-1\right)r^2dr}{e^{\beta\varepsilon_{\alpha\beta}}-1}+1\right)^{1/3}.$$

An additional 20 ns simulation was performed to determine the patch width and the width of the torsion angle between two particles for each patch. For these simulations, one protein copy was fixed in place by restraining the same set of alpha carbons. The harmonic spring was active with the same strength, but the pull rate was set to zero and the center of mass distance was kept at the equilibrium distance. Snapshots saved every 10 ps were analyzed to determine the vector perpendicular to the protein-protein interface. The dot product of the average vector with instantaneous vectors were calculated, and their mean was chosen as $\cos\delta_{\alpha}$, the patch width. The width of the dihedral angles, $\Delta\phi_{\alpha\beta}$, is calculated from the same trajectory by applying the rotation of the protein copies to a reference vector and calculating the angle between the two planes defined by (i) the first rotated vector and the center of mass vector, and (ii) the second rotated vector and the center of mass vector (fig. 2).

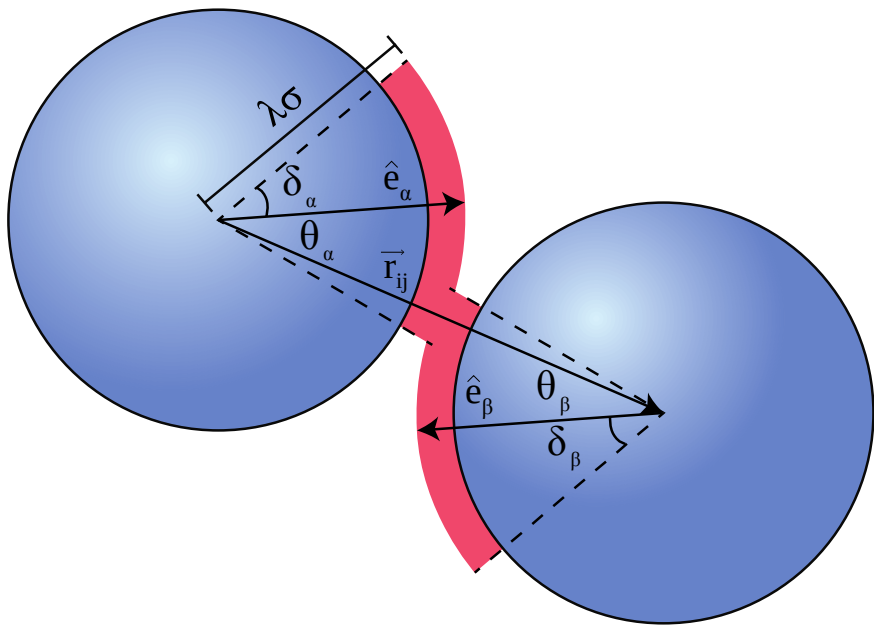

*Figure 1 Sample pair of interacting patches (red) illustrating the various model associated parameters $\delta_{\alpha}$, $\delta_{\beta}$, and $\lambda_{\alpha\beta}$. Here, $\widehat{e_{\alpha}}$ and $\widehat{e_{\beta}}$ point to the centers of patches α and β, respectively. If the particles are separated by less than λσ, they interact only if $\widehat{e_{\alpha}} \cdot \hat{r}_{ij} = \cos\theta_{\alpha} \geq \cos\delta_{\alpha}$ AND $\widehat{e_{\beta}} \cdot \widehat{-r_{ij}} = \cos\theta_{\beta} \geq \cos\delta_{\beta}$*

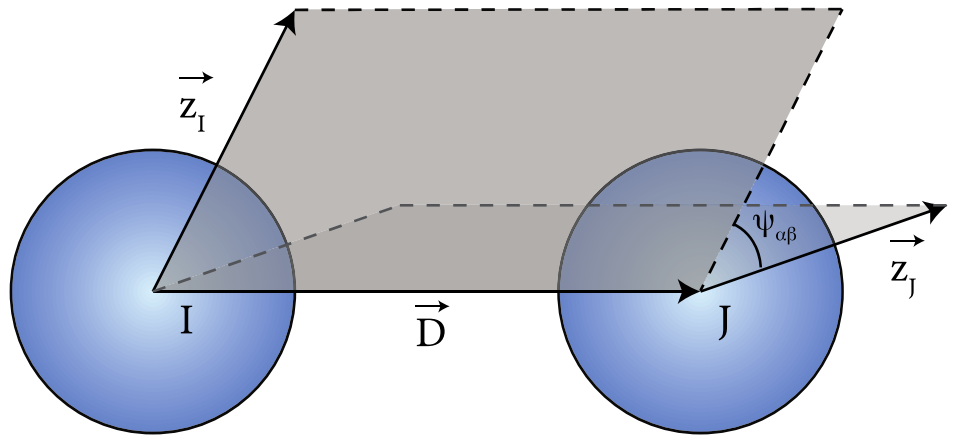

*Figure 2 Two overlapping patches interact when their orientation falls within the range of allowed dihedral angles. The dihedral, $\psi_{\alpha\beta}$, is the dihedral angle between vectors $\mathbf{z_I}$, $\mathbf{D}$, and $\mathbf{z_J}$, where $\mathbf{z_I}$ is calculated by applying the rotation of particle I to a reference axis chosen such that it is not parallel to any patch vector.*

The resulting parameters that result from this procedure are as reported in table S5 for contacts labeled cY, where Y is a Roman numeral for DBN contacts, and an Arabic numeral for DBI contacts. Contact 2 (c2) of DBI and contact I (cI) of DBN were found to be very similar in terms of geometry and the nature of their interactions and were hence merged. Similarly, contact V (cV) and VII (cVII) of DBN are almost identical and were also merged. The widths, range of dihedral angles, and interaction ranges for some patches for DBN had to be slightly widened to accommodate the crystal geometry of the model (see Model Geometry section below for details).

The model parameters obtained by all-atom simulations of the crystal structures give rise to patch free energies slightly higher than typical for protein crystal contacts, resulting in crystals remaining stable at unusually high temperatures. These patch free energies are nevertheless consistent with contact dissociation free energy estimates from the PISA server, it is therefore possible that some of the surface regions not considered as patches might have a slightly higher repulsion than estimated by the hard-core description. Because the qualitative picture that emerges from this analysis is only affected by relative patch strengths, however, this correction is not qualitatively significant. To correct for this effect, the patch deactivation temperature, $T_a$=1.9, is thus chosen such that the minimum packing fraction of DBI in its solubility line is around $10^{-4}$, as in experiments. (Note that the unit of temperature is such that T=1 corresponds to 277K, the experimental crystallization temperature for DBN.) We thus choose τ=0.05 so that patch deactivation occurs over a temperature range of order 10K, again paralleling the experimental observation. The triple point temperature obtained from simulations is $T_{tp}$=1.89, slightly below $T_a$, and is used in the main text to correct at once for the slight overestimate of the patch energies compared with the experimental values.

*Table S4 MD simulation parameters common to all types of all-atom simulations performed with Gromacs for model parameter determination.*

| |
|---|
| **Forcefield:** Amber03 [3] |
| **Water model:** TIP4P-EW [4] |
| **Temperature:** 277K for DBN, 310K for DBI |
| **Temperature control:** Nose-Hoover thermostat [5] |
| **Electrostatic method:** PME [6] |
| **Coulomb radius:** 1.4 nm |
| **van der Waals method:** cut-off |
| **van der Waals radius:** 1.4 nm |
| **Integration step:** 2 fs |
| **Constraints:** all bonds with Lincs [7] |
| **Energy minimization:** steepest-descent |
| **Salt concentration:** 160 mM of NaCl |
| **Pressure control:** Parrinello-Rahman barostat |

## Model Geometry

A number of modifications were made to the crystal geometry and to the patchy model so as to accommodate the relevant crystal forms and to simplify the computations. Because the DBI structure unit cell is almost orthorhombic: α=90°, β=91.29°, γ=90° we approximated it as a purely orthorhombic crystal. This results in minor changes in patch locations, which were previously shown not to affect phase diagram topology [8]. The other key feature is that the protein is slightly ellipsoidal, while a patchy particle is perfectly spherical. For DBI, all protein copies in the crystal are aligned, hence shrinking the unit cell along the *z*-direction solves the problem. For DBN, however, the structure is more complicated because the long axis of some of the eight asymmetric unit cells are aligned either in the *x*- or in the *y*-direction, thus requiring a more significant compression of the crystal. As a result, both the protein crystal and the patch positions must be perturbed to remove clashes. In addition, contacts II, cIV, and cVI must be collocated, in order to satisfy the various bonding constraints. The final geometry of both crystal models is summarized in tables S6 and S7. Interestingly, dihedral constraints are here necessary to properly separate the crystal ground states within the energy landscape.

*Table S5 Patch parameters obtained from all-atom molecular dynamics simulations. Numbers in parentheses denote the modified values to adjust the crystal geometry.*

| Contact | $\cos\delta_\alpha$ | $\cos\delta_\beta$ | $\Delta\phi_{\alpha\beta}$ (rad) | $\varepsilon$ ($k_BT$) | $\lambda$ ($\sigma$) |
|---|---|---|---|---|---|
| c1 | 0.956 | 0.963 | 0.07 | 21.5 | 1.025 |
| c2 | 0.972 (0.935) | 0.984 (0.935) | 0.21 | 6.3 | 1.1 |
| c3 | 0.992 | 0.994 | 0.14 | 10.3 | 1.06 |
| c4 | 0.997 | 0.997 | 0.06 | 15.8 | 1.025 |
| c5 | 0.947 | 0.966 | 0.13 | 6.0 | 1.037 |
| cI | 0.972 | 0.981 | 0.21 | 4.9 (6.3) | 1.06 (1.1) |
| cII | 0.998 (0.96) | 0.999 (0.96) | 0.06 (0.3) | 8.7 | 1.05 (1.11) |
| cIII | 0.998 (0.95) | 0.997 (0.95) | 0.038 (0.2) | 10.1 | 1.03 (1.1) |
| cIV | 0.970 (0.96) | 0.937 | 0.24 (0.3) | 4.3 | 1.14 |
| cV | 0.995 (0.95) | 0.989 (0.95) | 0.10 (0.21) | 18.8 (18.5) | 1.06 (1.153) |
| cVI | 0.898 | 0.927 | 0.23 (0.3) | 8.6 | 1.03 (1.11) |
| cVII | 0.991 (0.95) | 0.998 (0.95) | 0.10 (0.21) | 18.2 (18.5) | 1.06 (1.53) |

*Table S6 Patch locations and dihedral angles. Because of geometric constraints of the DBN structure, contacts cII, cIV, and cVI are collocated.*

| Contact | $\hat{e}_1$ | $\hat{e}_2$ | $\hat{e}_3$ | $\phi_{\alpha\beta}$ |
|---|---|---|---|---|
| c1 | 0.6614<br>0.6473 | 0.6543<br>-0.2853 | 0.3667<br>0.7068 | -2.2703 |
| c2/cI | -0.9854<br>0.9854 | 0.0471<br>-0.0471 | 0.1101<br>-0.1101 | 0.0 |
| c3 | -0.0141<br>0.0141 | -0.9402<br>0.9402 | 0.3403<br>-0.3403 | 0.0 |
| c4 | 0.3231<br>0.3089 | 0.1911<br>-0.7486 | -0.9269<br>-0.5867 | -2.1288 |
| c5 | -0.6474<br>-0.6613 | 0.2851<br>-0.6543 | -0.7068<br>-0.3669 | 2.2698 |
| cII* | -0.9494<br>0.0011 | 0.3101<br>0.4334 | -0.0499<br>0.9012 | -2.4946 |
| cIII | -0.6455<br>-0.7102 | -0.5123<br>-0.5524 | -0.5665<br>0.4365 | -1.80 |
| cIV* | -0.9494<br>0.0011 | 0.3101<br>0.4334 | -0.0499<br>0.9012 | -2.4946 |
| cV/VII | 0.3162<br>0.4927 | -0.8026<br>0.8531 | -0.5058<br>0.1716 | -2.1529 |
| cVI* | -0.9494<br>0.0011 | 0.3101<br>0.4334 | -0.0499<br>0.9012 | -2.4946 |

*Table S7 Geometry of DBI and DBN crystals for the patchy model. Lengths are in units of σ. The rotation of each particle is calculated from Euler angles reported as (α, β, γ) with the $R_z(\alpha)R_x(\beta)R_z(\gamma)$ convention, where $R_z$ denotes a counterclockwise rotation through the z-axis and $R_x$ a counterclockwise rotation through the x-axis.*

| DBI | 2 asymmetric units, unit cell: $(\sqrt{2}, 1, 1)$ |
|---|---|

| | | |
|---|---|---|
| | Particle position | Particle rotation |
| | 0, 0, 0 | -1.927, 1.349, -1.474 |
| | $\sqrt{2}/2, 0.5, 0.5$ | 1.927, 1.793, 1.668 |
| DBN | 8 asymmetric units, unit cell: (1.02, 3.149,3.586) | |
| | Particle position | Particle rotation |
| | 0.184, 0.470, 0.385 | 1.427, 1.681, 1.371 |
| | 0.183, 0.073, 3.061 | 1.096, 0.186, 5.375 |
| | 0.761, 1.024, 3.238 | 1.685, 1.547, 4.421 |
| | 0.647, 1.438, 0.631 | 2.043, 2.916, 2.222 |
| | 0.954, 2.030, 1.478 | 4.869, 1.420, 4.519 |
| | 0.032, 1.704, 2.446 | 4.876, 3.029, 1.923 |
| | 0.479, 2.623, 2.213 | 4.548, 1.581, 1.316 |
| | 0.537, -0.203, 1.167 | 3.468, 0.065, 6.046 |

## Phase Diagram Determination

The phase diagram of the patchy particle model is computed by specialized MC simulations. In general, 200,000 MC sweeps each of which consists of N displacement and N rotational moves for constant NVT simulation and an additional two volume moves for NPT simulations. The amplitude of the displacements is chosen such that an acceptance rate around 50% is obtained. System sizes are chosen to be similar from one phase to the other, while respecting the crystal symmetry, here, $N_{DBI} = 432$ and $N_{DBN} = 512$. For convenience, these simulations report distances in units of $\sigma$, and energies in units of $k_B T$ with $T = 277K$.

Crystal free energies were obtained by integrating from an ideal Einstein crystal, and then integrated along isobars to determine the free energy as a function of temperature. Thermodynamic integration along isobars is more numerically stable than over integration along isotherms because both crystals are strongly geometrically constrained, and thus a very low compressibility. For a system with temperature-dependent pair potential, the chemical potential of the system with respect to a reference system at $T_o$ is given by

$$\beta\mu(T,p) = \beta_0\mu(T_o,p) - \int_{T_0}^{T} \frac{H(T')}{Nk_B T'^2} dT' + \int_{T_0}^{T} \frac{< \frac{\partial U}{\partial T'} >}{Nk_B T'} dT',$$

where $H$ is the enthalpy, and $< \frac{\partial U}{\partial T} >$ is the ensemble average of the temperature derivative of the pair potential.

Sampling the low density–low pressure regime of the phase diagram is numerically inefficient, hence in this regime, we estimate the fluid equation of state and free energy from the second virial coefficient:

$$\frac{\beta p}{\rho} = 1 + B_2\rho,$$
$$\beta\mu = \beta\mu^{id} + 2B_2\rho = \log \Lambda^3\rho + 2B_{22}\rho,$$

where $\mu^{id}$ is the chemical potential of the ideal gas, and the thermal de Broglie wavelength $\Lambda$ is set to one without loss of generality.

Coexistence points were obtained from the intersection of the crystal and fluid isobaric temperature-chemical potential curves. Solubility lines were then traced by numerical Gibbs-Duhem integration [9] [10]. Both boxes were simulated for 200,000 MC sweeps for each temperature. The slope of the solubility line was corrected every 1000 MC sweeps. The temperature was propagated in steps of $\Delta\beta = 0.003$. Note that introducing temperature-dependent patches requires modifying the standard Clausius-Clapeyron equation [11]

$$\frac{dP}{dT} = \frac{\Delta e + p\Delta v - T\Delta \langle \frac{\partial u}{\partial T} \rangle}{T\Delta v},$$

where $\frac{\partial u}{\partial T}$ is the derivative of the interaction energy with respect to temperature, the angular brackets denote an ensemble average. This additional term here allows for the crystal solubility curve to be inverted. (Typically, $dP/_{dT}$ is positive because $\Delta e$ and $\Delta v$ tend to have the same sign.) At low fluid densities, only the crystal was simulated, while $e_f$ was calculated analytically

$$e_f = \frac{1}{N}\left(\frac{d(\frac{F}{T})}{d(\frac{1}{T})}\right)_{N,V} = \rho \left(\frac{\partial B_{22}}{\partial \beta}\right)_{N,V}.$$

The ensemble average of the interaction energy can be estimated by considering the average number of bonds active in the fluid, $e_f/e_{tot}$, where the denominator is the sum of all patch energies

$$\langle \frac{\partial u}{\partial T} \rangle_f = \frac{\rho}{e_{tot}}\left(\frac{\partial B_{22}}{\partial \beta}\right)_{N,V} \frac{\partial u}{\partial T}.$$

The second virial coefficient for our model is calculated as in Ref. [12] and is defined as

$$B_{22} = -\frac{1}{2V}\int d\boldsymbol{r_1} \int d\boldsymbol{r_2} \int d\boldsymbol{\Omega_1} \int d\boldsymbol{\Omega_2} \int d\chi \left(e^{-\beta u(|\boldsymbol{r_1} - \boldsymbol{r_2}|)} - 1\right),$$

where the integration is performed over the particle positions, $\boldsymbol{r}$, orientation (in terms of solid angles, $\boldsymbol{\Omega}$), and dihedral angle $\chi$ (see figure 2). Because cII, cIV, and cVI overlap in our patchy model, but do not have the same ranges or angular widths, we have

$$\frac{B_{22}}{B_{22}^{HS}} = 1 - \sum (\lambda_{ij}{}^3 - 1)\sin^2\left(\frac{\delta_i}{2}\right)\sin^2\left(\frac{\delta_j}{2}\right)\left(e^{\beta\varepsilon_{ij}} - 1\right)(2\Delta\phi_{ij})$$
$$- (\lambda_{IV}^3 - \lambda_{II}^3)\left(e^{\beta\varepsilon_{IV}} - 1\right)\sin^2\left(\frac{\delta_{IV,\alpha}}{2}\right)\sin^2\left(\frac{\delta_{IV,\beta}}{2}\right)(2\Delta\chi) - (\lambda_{II}^3 - 1)\left(e^{\beta(\varepsilon_{II}+\varepsilon_{IV}+\varepsilon_{VI})}\right.$$
$$\left. - 1\right)\sin^2\left(\frac{\delta_{II,\alpha}}{2}\right)\sin^2\left(\frac{\delta_{II,\beta}}{2}\right)(2\Delta\chi) - (\lambda_{VI}^3 - 1)\left(e^{\beta(\varepsilon_{IV}+\varepsilon_{VI})}\right.$$
$$\left. - 1\right)\sin^2\left(\frac{\delta_{IV,\alpha}}{2}\right)\sin\left(\frac{\delta_{IV,\beta} + \delta_{IV,\alpha}}{2}\right)\sin\left(\frac{\delta_{IV,\beta} - \delta_{IV,\alpha}}{2}\right)(2\Delta\chi)$$
$$- (\lambda_{VI}^3 - 1)\left(e^{\beta\varepsilon_{VI}}\right.$$
$$\left. - 1\right)(2\Delta\chi)\left[\sin^2\left(\frac{\delta_{IV,\alpha}}{2}\right)\sin\left(\frac{\delta_{IV,\alpha} + \delta_{IV,\beta}}{2}\right)\sin\left(\frac{\delta_{IV,\alpha} - \delta_{IV,\beta}}{2}\right) + \sin\left(\frac{\delta_{VI,\alpha} - \delta_{IV,\alpha}}{2}\right)\sin\left(\frac{\delta_{VI,\alpha} + \delta_{IV,\alpha}}{2}\right)\sin^2\left(\frac{\delta_{VI,\beta}}{2}\right)\right]$$

where $2\Delta\chi = 0.6$, the width of the range of valid dihedral angles, is the same for all three patches.